\documentstyle[12pt]{article}
\newenvironment%
     {abst}%
     {\normalsize \begin{center}%
     {ABSTRACT} \end{center}
     \baselineskip 12pt
     \leftmargin 3pc
     \rightmargin 3pc
    \quotation}%
     {\endquotation}%
\newcommand{\be}{\begin{equation}}
\newcommand{\ee}{\end{equation}}
\newcommand{\bea}{\begin{eqnarray}}
\newcommand{\eea}{\end{eqnarray}}
\newcommand{\vac}{\left.\mid 0\right>}
\newcommand{\k}{{\bf k}}
\newcommand{\q}{{\bf q}}
\newcommand{\x}{{\bf x}}
\newcommand{\y}{{\bf y}}

\newcommand{\n}{\bigtriangledown}
\begin{document}
\begin{flushright}
{\bf ISU-IAP.Th94-03,\\
 Irkutsk}
\end{flushright}
\vspace{2cm}

\begin{center}
{\large \bf ANALYSIS OF ONE PARTICLE EXCITATIONS \\
     IN PHENOMENOLOGICAL MODELS OF QCD}\footnote{
This work is partially supported by RFFI
N 938-18-35
and by grant of the natural history
fundamentals (St-P.) N 94-6.7-2057.}

\vskip 1cm
{A.N.VALL, S.I.KORENBLIT, V.M.LEVIANT, A.V.SINITSKAYA}
\vskip .5cm
{\footnotesize {\em Department of Theoretical Physics,\\
 Irkutsk State University, \\
 Irkutsk, 664003, RUSSIA.}\\
e-mail: VALL@physdep.irkutsk.su}
\end{center}

\begin{abst}
In this paper for the quark Nambu-Jona-Lasinio model and for
the phenomenological four fermionic model of QCD
we analyze vacuum energy structure and calculate dependence of energy
of one-particle excitations on momentum. We reveal existence of
two different types of excitations above the vacuum with the respective
minimization of the vacuum energy density. First type of excitation energy
spectrum is the usual spectrum of relativistic massive particle, but the
second one has a linear dependence on momentum.
\end{abst}
\newpage
\vskip -3mm

\begin{flushleft}
\bf 1. Introduction
\end{flushleft}
\vskip 1mm

In their pioneer paper$^1$ Nambu and Jona-Lasinio (NJL),
using the analogy with superconducive mechanism, had shown an
intimate connection of unitary\--\-nonequivalent transformations
of quantum states with the existence of nontrivial vacuum.
Later this approach was generalized on a case of quark
interaction in chiral QCD limit (see, e.g., review$^2$).
In this paper we consider one particle excitations
of vacuum and its energy spectrums for wide set of permissible
transformations. We show that besides the known NJL solution
there always exists one more excitation with a linear spectrum
on momentum and this fact is a property of any contact four -fermionic
$SU(2)\times SU(2)\times U(1)$-symmetric interaction.
\vskip .8cm

\begin{flushleft}
\bf 2. Analysis of the NJL solution.
\end{flushleft}

The NJL Hamiltonian is
\be
  H = \int d^3x\bar{\psi}(x)\hat{\vec{p}}\vec{\gamma}\psi (x) -
     \frac{G_1}{2}\int d^3x\left[\left(\bar{\psi}(x)\psi (x)\right)^2 -
     \left(\bar{\psi}(x)\gamma^5\vec{\tau}\psi (x)\right)^2\right],
\label{1}
\ee
where $\hat{\vec{p}} = - i\vec{\n}$,
color and flavor indexes on $\psi$ are assumed.
The NJL transformation representing Heisenberg field
$\psi (x)$  at $t=0$ in terms of massless spinors $u(p),\;v(p)$ leads to the
following form of the Heisenberg field (dynamical mapping$^{3}$)
\bea
\psi ({\bf x},0) &=&
\frac{1}{(2\pi)^{\frac{3}{2}}}\int d^3p\left[N^r(p)a^r(p)
e^{i{\bf p}{\bf x}}
 + M^r(p)b^{+r}(p)e^{-i{\bf p}{\bf x}}\right] \label{2}\\
N^r(p) &=& \left(\cos\theta(p)  +  \gamma^0\sin\theta(p)\right)u^r(p)
\nonumber \\
M^r(p) &=& \left(\cos\theta(p)  -  \gamma^0\sin\theta(p)\right)v^r(p),
\nonumber
\eea
$u^r(p),\; v^r(p)$
are the massless spinors ($\rlap/p u(p) =
\rlap/p v(p) = 0$) normalized by the usual conditions
$u^{+r}(p)u^s(p)~ = v^{+r}(p)v^s(p) = \delta^{rs} $. It is a simple
exercise to check that the spinors $N^r(p)$ and $M^r(p)$ satisfy the
equations:
\bea \left(E(p)\gamma^0 - \vec{p}\vec{\gamma} -
m^*(p)\right)N^r(p) &=& 0 \nonumber  \\
\left(E(p)\gamma^0 -
\vec{p}\vec{\gamma} + m^*(p)\right)M^r(p) &=& 0, \label{3}
\eea
with $E(p) = p/\cos2\theta(p),\;\;\; m^*(p) = p\tan
2\theta(p),\;\;\;p = \mid{\bf p}\mid$.  Creation and annihilation
operators $a^r(p),\;b^r(p)$ and $a^{+r}(p),\; b^{+r}(p)$ satisfy the
conventional anticommutation relations for fermions and are
interpreted in a sense of dynamical mappings as operators of
excitations of "physical" particles. Vacuum state is defined
with respect to them.

Vacuum energy
$W_0 = \left< 0\mid\right.H\vac$ evaluated by using
Eqs.(\ref{2})
is turned out to be a functional of the transformation angles $\theta(p)$:
\bea
\frac{W_0}{V} &=& -2N_cN_f\frac{1}{(2\pi)^3}\int
k\cos(2\theta(k))d^3k  - \nonumber \\
&-& N_cN_fG\left[\frac{1}{(2\pi)^3}\int
\sin(2\theta(k))d^3k\right]^2 - \frac{2G_1N_cN_f}{(V^*)^2},
\label{4}
\eea
where $G = G_1(1 + 2N_cN_f),\;\; \frac{1}{V^*} =
\frac{1}{(2\pi)^3}\int d^3k,\;\; V = \int d^3x$.
Thus, $\theta(p)$ "enumerates" all possible vacuums.
Variation of $\theta(p)$ gives
the following condition for the extremum:
\bea 1 &=&
\frac{G}{(2\pi)^3}\int d^3k\frac{1}{\sqrt{{\bf k}^2 +
m^{*2}}},\;\;\;m^*\not=0 \label{5} \\
m^* &=& p\tan 2\theta(p) = const.
\nonumber
\eea
Another trivial solution corresponds to $m^* = 0$.
To answer the question whether the solution suits for maximum
or for minimum of $W_0$ we need to check the sign of functional determinant:
\be
\det\left[\frac{\delta^2W_0}{\delta\theta(k)\delta\theta(p)}\right]_F =
N\left[1 - \frac{2G}{(2\pi)^3}\int d^3k\frac{k^2}{E(k)^3}\right] =
N\frac{G}{(2\pi)^3}\int d^3k \frac{m^2}{E(k)^3} > 0. \label{6}
\ee
$$
N = e^{\int d^3k\ln(8N_cN_fE(k))}.
$$
So, if $m^* \not = 0$ this extremum is minimum which is degenerated
at $m^* = 0$ into the inflection point.

The energy spectrum of one-particle excitation $E^*(k)$ is determined by
the following eigenvalue equation:
\be
\left[H, d^{+r}(k)\right] \vac = E^*(k)d^{+r}(k)\vac,  \label{7}
\ee
$d^{+r}(k)$ stands for $a^{+r}(k)$ or $b^{+r}(k)$.
Calculating the commutator in the l.h.s. of above equation using the
dynamical mapping (\ref{2}) and omitting three-particle excitations
we obtain expression for $E^*(k)$:
\be
E^*(k) = k\cos(2\theta(k)) + \frac{G}{V^*}\sin^2(2\theta(k)). \label{8a}
\ee
On the solution (\ref{5}) $E^*(k)$ coincides with $E(k)$
\be
E^*(k) = \sqrt{{\bf k}^2 + m^{*2}} = E(k). \label{8}
\ee

Now we want to draw attention to the important fact, that under the
functional variation of $W_0$ over $\theta(k)$ one more solution is lost,
and it should be studied separately. Namely, the solution:
$\theta(k) = \phi = const$. This solution, as will be shown below,
admits an extremum in $W_0$, and this extremum corresponds to minimum
of $W_0$.
In this case $m^*$ is a linear function of momentum: $m^*(k) = ak$,
and the expression for the vacuum energy has the form:
\bea
\frac{W_0}{V} &=& -2N_cN_f\left[\frac{<k>}{V^*}\cos 2\phi +
\frac{G}{2V^{*2}}\sin^22\phi\right] -\frac{2G_1N_cN_f}{V^{*2}} \label{9} \\
<k> &=& \frac{\int kd^3k}{\int d^3k}. \nonumber
\eea
 From (\ref{9}) we obtain:
\bea
\frac{d}{d\phi}\frac{W_0}{V} &=& 4N_cN_f\frac{<k>}{V^*}\sin 2\phi
\left(1 - \frac{G}{<k>V^*}\cos 2\phi\right) = 0 \label{10} \\
\frac{d^2}{d\phi^2}\frac{W_0}{V} &=& 16N_cN_f\frac{G}{V^{*2}}
\sin^2 2\phi > 0. \label{11}
\eea
Eq.(\ref{10}) determines the
parameter $\phi$ in the extremum point:
\be
\cos 2\phi = \frac{<k>V^*}{G} \equiv c^*, \label{12}
\ee
and this extremum is in fact always minimum (Eq.(\ref{11})). Hereafter
to the end of the text we will deal with positive values of $c^*$, so
that the minimum existence condition is just restriction $c^* < 1$
(recall, that we use  the system $c = \hbar = 1$) and this means $c^*$
is less than speed of light.
Inserting the expression of $\cos2\phi$ via $c^*$ (Eq.(\ref{12})) at
$\theta(k) = \phi$ into Eq.(\ref{8a}) we finally obtain
the energy spectrum $E^*$ of excitation
(\ref{7}):
\be
E^*(k) = c^*k + \frac{G}{V^*}\left(1 - c^{*2}\right), \label{13}
\ee
As one can recognize from Eq.(\ref{13}), $c^*$ has the meaning
of group velocity of the excitation.
Thus, the spectrum is proved to be a linear function of momentum , i.e.
"neutrino" type, and the condition of the minimum existence corresponds
to the demand $c^* < 1$. At $c^* = 1$ the minimum of $W_0$ is degenerated
into the inflection point.

So, NJL transformations for the Hamiltonian (\ref{1}) yield to
two types of stable excitations above vacuum. One excitation describes a
massive particle with the normal relativistic spectrum, and the mass
arising dynamically. The second excitation has the linear energy
spectrum.
This fact is a consequence of
unitary-nonequivalenceness of the transformations (\ref{2}) from the
massless particles to the massive ones causing the vacuum to be
nontrivial.
As will be shown later appearing of these two
types of excitations is not a result of the given transformation
(\ref{2}) and of the Hamiltonian choice (\ref{1}). They arise
generally for any contact four-fermionic $SU(2)\times SU(2) \times
U(1)$- globally symmetric interaction. Let us note, eventually,
that the NJL transformations can be represented in terms of quantum
fields as:
$$
\psi({\bf x},0) = e^{\hat{G}}q({\bf x},0)e^{-\hat{G}}
$$
\vskip -3mm
\be
G = \int d^3xd^3y\left[ \bar{q}^{(-)}({\bf x},0)\ell({\bf x} -
{\bf y})q^{(-)}({\bf x},0)\right. -
\left.\bar{q}^{(+)}({\bf x},0)\ell({\bf x} -
{\bf y})q^{(+)}({\bf x},0)\right], \label{14}
\ee
where $q(x) = q^{(+)}(x) + q^{(-)}(x)$ is massless spinor
field expanded on its frequency parts, and $\theta(k)$ is Fourier
image of the real function $\ell({\bf x})$.
\vskip 1cm

\begin{flushleft}
\bf 3. Model of pseudo-scalar couplings.
\end{flushleft}

Consider now $SU(2)\times SU(2)\times U(1)$-invariant
Lagrangian$^2$:
\bea
L = \bar{\psi}i\rlap/\partial\psi + &\frac{G_1}{2}&
     \left[\left(\bar{\psi}(x)\psi (x)\right)^2 +
     \left(\bar{\psi}(x)i\gamma^5\vec{\tau}\psi (x)\right)^2\right] -
\nonumber \\
-&\frac{G_2}{2}&
     \left[\left(\bar{\psi}(x)\vec{\tau}\gamma_\mu\psi (x)\right)^2 +
     \left(\bar{\psi}(x)\gamma^5\gamma_\mu\vec{\tau}\psi(x)
     \right)^2\right] - \nonumber \\
-&\frac{G_3}{2}&
      \left[\left(\bar{\psi}(x)\sigma_{\mu\nu}\psi (x)\right)^2 +
     \left(\bar{\psi}(x)i\sigma_{\mu\nu}\gamma^5\vec{\tau}\psi
     (x)\right)^2\right]. \label{15}
\eea
Our aim is to quantize the respective Hamiltonian by the
dynamical mapping method$^3$ in such a way that the relevant vacuum
would contain condensate with the quantum numbers of $\pi$-mesons.
Therefore, we take generator of the transformations from "physical"
fields to the Heisenberg ones (dynamical mapping) exploiting
the expression for the axial charge.
\be
\psi({\bf x},0) = e^{\hat{G}}q_p({\bf x},0)e^{-\hat{G}} \label{16}
\ee
where $q_p({\bf x},0)$ is a "trial" physical massive field with a
mass $m(k)$ arbitrarily depending on momentum.
\bea
q^{(+)}({\bf x},0) &=&\frac{1}{(2\pi)^{\frac{3}{2}}}\sum_r\int d^3k g(k)
\frac{m(k)}{E(k)}u^r(k)a^r(k)e^{i{\bf k}{\bf x}}, \nonumber \\
q^{(-)}({\bf x},0) &=&\frac{1}{(2\pi)^{\frac{3}{2}}}\sum_r\int d^3k
\bar{g}(k)\frac{m(k)}{E(k)}v^r(k)b^{+r}(k)e^{-i{\bf k}{\bf x}}, \label{17}
\eea
$u^r(k)$ and $v^r(k)$ are massive spinors normalized by the conditions
$\bar{u}^r(k)u^s(k) = - \bar{v}^r(k)v^s(k) = \delta^{rs}, \;\;\;E(k) =
\sqrt{{\bf k}^2 + m^2(k)}$. Operators $a^r$ and $b^r$ anticommute as:
\be
\left\{b^r(k),\;b^{+s}(p)\right\} = \left\{a^r(k),\;a^{+s}(p)\right\} =
\frac{E(k)}{m(k)}\delta^{rs}\delta({\bf k} - {\bf p})I_cI_f, \label{18}
\ee
$I_c$ and $I_f $ are unit flavor and color matrixes
(tr$I_\alpha = N_\alpha$). Wave function
$g(k)$ is introduced for "smooth" regularization of divergent integrals
and to attach physical meaning to them. If quantums of physical fields are
interpreted as in paper$^3$, then $g(k)$  characterizes a momentum
distribution inside the excitation, and the integral $\frac{1}{(2\pi)^3}
\int \mid g(k)\mid^2  d^3k = \frac{1}{V^*}$ determines $V^*$ - space volume
of this excitation. We normalize $ \mid g(k)\mid^2 $ in such a manner
that it is $\sim 1$ in the region where it essentially differs from zero,
i.e. we choose:
\be
\left<\mid g(k)\mid^2\right>_g = \frac{\int \mid g(k)\mid^4 d^3k}{
\int \mid g(k)\mid^2d^3k} = 1  \label{19}
\ee
Correspondence with the usual momentum "cut-off" parameter $\Lambda$ is
established due to the equality:
$$
\frac{1}{(2\pi)^3}\int^\Lambda d^3k = \frac{1}{V^*}.
$$
So, in the local (over momentum) relations  we will put $\mid g(k)\mid^2 =
1$ but saving it in the integrals to regulate its convergence. Taking into
account all above comments we construct the generator $\hat{G}$
from the diagonal (with respect to the frequencies) parts of the
axial charge:
\be
\hat{G} = i\int d^3x \left[\bar{q}^{(-)}({\bf x},0)\gamma^0\gamma^5
\hat{\tau}q^{(-)}({\bf x},0)
- \bar{q}^{(+)}({\bf x},0)\gamma^0\gamma^5
\hat{\tau}q^{(+)}({\bf x},0)\right],
\label{20}
\ee
where $\hat{\tau} = \ell\cdot (\vec{n}\vec{\tau}),\;\;
\ell = const,\;\;
\vec{n}$ is a unit vector in the flavor space. With such choice of $G$
we obtain the following dynamical mapping:
\bea
\psi (0,{\bf x}) &=&
\frac{1}{(2\pi)^{\frac{3}{2}}}\sum_r\int d^3k\frac{m(k)}{E(k)}
\left[g(k)N^r(k)a^r(k)e^{i{\bf k}{\bf x}}
 + \bar{g}(k)M^r(k)b^{+r}(k)e^{-i{\bf k}{\bf x}}\right] \nonumber \\
N^r(p) &=& \left[\cos(\ell\Phi(k))  + i\vec{n}\vec{\tau}
\gamma^0\gamma^5\sin(\ell\Phi(k))\right]u^r(p)
        \nonumber     \\
M^r(p) &=& \left[\cos(\ell\Phi(k))  - i\vec{n}\vec{\tau}
\gamma^0\gamma^5\sin(\ell\Phi(k))\right]v^r(p)
\label{21}
\eea
 From above it follows that
$\ell$ and $\Phi(k) = \frac{m(k)}{E(k)}$ play the role of parameters
of $\hat{G}$-trans\-for\-mation, i.e. as in Eq.(\ref{4}) they "enumerate"
all possible unitary-nonequivalent vacuums. For the vacuum energy
density we find out the expression:
\bea
\frac{\left<0\mid\right.H\vac}{V} =
&=& -2N_cN_f\left\{\frac{1}{(2\pi)^3}\int d^3k
\mid g(k)\mid^2\cdot \frac{{\k}^2}{E(k)}\cos(2\ell\Phi(k))\right. +
\nonumber\\
&+&\frac{G}{2}\left[\frac{1}{(2\pi)^3}\int d^3k
\mid g(k)\mid^2\cdot \Phi(k)\cos(2\ell\Phi(k))\right]^2 + \nonumber \\
&+& \left.\frac{G}{2}\left[\frac{1}{(2\pi)^3}\int d^3k
\mid g(k)\mid^2\cdot\sin(2\ell\Phi(k))\right]^2\right\} +
const, \label{22}
\eea
where $G = G_1 + 2G_1N_cN_f + 2G_3$, and the $const$ does not depend
on $\ell$ and $\Phi(k)$. Hence it follows, firstly, that the terms of the
Lagrangian (\ref{15}) corresponding to the vector and axial-vector hadronic
currents ($\rho$ and $A_1$ mesons) do not contribute to the vacuum energy
density. Secondly, each $SU(2)\times SU(2)\times U(1)$-invariant term of
the Lagrangian (\ref{15}) gives the uniform contribution to $W_0$, so that
accounting all the terms of (\ref{15}) results in the summary constant $G$.
Let us vary $W_0$ on $\Phi(k)$, reckoning the physical vacuum to be
a minimum of the energy density $W_0/V$, and equating the variation to zero.
Using the relation $\delta (\k^2/E(k)) = -m(k)\delta\Phi(k)$ we
have an integral nonlinear equation for $m(k)$
\bea
&&\left(\frac{\k^2}{E(k)} + \mu_1\cdot\Phi(k)\right)\tan(2\ell\Phi(k)) =
\mu_2 + \frac{\mu_1 - m(k)}{2\ell} \label{23} \\
&&\mu_1 = \frac{G}{(2\pi)^3}\int d^3k \mid g(k)\mid^2\cdot\Phi(k)
\cos(2\ell\Phi(k)); \nonumber \\
&&\mu_2 = \frac{G}{(2\pi)^3}\int d^3k
\mid g(k)\mid^2\cdot\sin(2\ell\Phi(k)). \nonumber
\eea
This equation should be true for the all admitted values of $k$:
$0 \leq k < \infty$.
The analysis shows that the equation (\ref{23}) has the only solution
$m(k) = ak$, with $a = const$ satisfying the transcendental equation:
\be
\frac{1}{\sqrt{1 + a^2}}\tan\left(\frac{2\ell a}{\sqrt{1 + a^2}}\right) =
- \frac{a}{2\ell}. \label{24}
\ee
This equation is thought to connect the parameter $\ell$
with the constant $a$.
There are many solutions of (\ref{24}) written as
\be
\frac{2\ell a}{\sqrt{1 + a^2}} = 2\pi\left(n + \frac{3}{4}\right)
\left(1 + \frac{1}{a^2} + \cdots \right),\;\;n = 0,1,2,\dots
\label{25}
\ee
The expression for $W_0/V$ under condition (\ref{24}) has the following
form:
\be
\frac{W_0}{V} = -2N_cN_f \frac{<k>_g}{V^*}\left(f(a) -
\frac{G}{2V^*<k>_g}f^2(a)\right) + const, \label{26}
\ee
denoting
\be
f(a) = \frac{\cos\left(\frac{2\ell a}{\sqrt{1 + a^2}}\right)}
{\sqrt{1 + a^2}}, \label{26b}
\ee
and
$$
<k>_g = \frac{\int k \mid g(k)\mid^2d^3k}{\int \mid g(k)\mid^2d^3k}.
$$
Now, in order to fix the value of $\ell$ (or $a$) we have
to minimize (\ref{26}) over $\ell  (a)$.
The very form of (\ref{26}) hints to take derivative with respect
to $a$ rather than $\ell$, so we do:
\bea
\frac{d}{da}\frac{W_0}{V} &=& -2N_cN_f\frac{<k>_g}{V^*}f'(a)\left(
1 - \frac{G}{V^*<k>_g}f(a)\right) = 0  \nonumber \\
\frac{d^2}{da^2}\frac{W_0}{V} &=& 2N_cN_f\frac{<k>_g}{V^*}
\left(f'(a)\right)^2 >0 \label{27}
\eea
Thus, as in above, the extremum exists and always corresponds to minimum.
First of Eqs.(\ref{27}) links $f(a)$ with $c^*$
\be
f(a) = \frac{V^*}{G}<k>_g \equiv c^*,    \label{28}
\ee
besides the expression for $c^*$  via the parameters
of the model at hand coincides exactly with
the expression obtained under the NJL transformations Eq.(\ref{12}),
while the second of Eqs.(\ref{27}) constrains $c^* < 1$.

Evaluation of excitation energy $E^*$ from Eq.(\ref{7}) results
(compare with (\ref{8a}))
\be
E^* = \frac{\k^2}{E(k)}\cos(2\ell\phi(k)) +
\mu_1\Phi(k)\cos(2\ell\Phi(k)) + \mu_2\sin(2\ell\Phi(k)) \label{29a}
\ee
What is most surprising, the Eq.(\ref{29a}) exactly coincides with
expression (\ref{13}) on the solution (\ref{28})
\be
E^*(k) = c^*k + \frac{G}{V^*}\left(1 - c^{*2}\right).
\label{29}
\ee

If in Eq.(\ref{21}) the parameter $\ell$ is chosen
to be a function of $k: \ell =\ell (k)$, that corresponds to the input
of a function $\tilde{\ell}(\x - \y)$ into the generator $\hat{G}$
Eq.(\ref{20}), then
it is not hard to make sure that the equation on the extremum of vacuum
energy in this case yields for $m(k)$  the NJL solution, i.e.
$m(k) = m^* = const.$

We had studied all possible generators $G$, composed from the bilinear
combinations of fields. From the results of calculations there follows the
only conclusion that in the models with the contact four fermionic
interaction possessing the $SU(2)\times SU(2)\times U(1)$ symmetry
two types of stable excitations emerge - with the universal expressions for
one particle energies Eq.(\ref{8}) or Eq.(\ref{13}) and with respective
(in terms of $m^*$ or $c^*$) universal expressions for vacuum energies.
The linear spectrum at $c^* = 1$ corresponds to the neutrino spectrum.
However, this point cannot be achieved because of two reasons.

First: at $c^* = 1$ the minimum degenerates into the inflection point.

Second: rewriting (on $m(k) = ak$ with $1/a = q_0$)
the expression for the spinor $u^r(k)$ in the form
\bea
u^r(k) &=& \frac{1}{\sqrt{2\left(1 + \sqrt{1 + q_0^2}\right)}}\cdot
\left(1 + \sqrt{1 + q^2_0} - \vec{q}_0\vec{\gamma}\right)u^r_0,
\label{30} \\
(\k &=& {\bf n}k,\;\;\q_0 = {\bf n}q_0) \nonumber
\eea
and finding out the following behavior of $q_0$ near $c^* \sim 1$
\be
q_0 \sim \frac{c^*}{\sqrt{1 - c^{*2}}}  \label{31}
\ee
one can see, owing to the relativistic factor in Eq.(\ref{31}),
the second obstacle to achieve the point $c^* = 1$.

Finally note, that the relativity - invariant generators $\hat{G}$,
constructed from all bilinear combinations of fields can be brought
to any of the forms $a^+b^+, ab, ect$. However, when the algebra of
the generators is observed
it is infinite\--\-dimensional
because of the relativistic phase volume
$d^3km(k)/E(k)$,
since every generator contains
different orders of $m(k)/E(k)$. The exception concerns the case of
linear spectrum  with $m(k)/E(k) = const$, then algebra is closed and
the parameter space becomes finite\--\-dimensional.

It is worth mentioning in a conclusion  that the energy spectrum
can be rewritten in the more convenient form for the experimental
treatment, namely:
\be
E^*(k) = \chi_{0} c k + \frac{G}{V^*}\left(1 - \chi_0^{2}\right).
\label{32}
\ee
Here $\chi_{0}$ is a dimensionless number, $c$ is the speed of light
in "emptiness", and the vacuum stability condition reduces in this
case to $\chi_{0} < 1$. Parameter $\frac{G}{V^*}$ has dimension of
energy and, as well as the parameter $\chi_{0}$, is not determined by the
initial Lagrangian. Its determination needs a model for the
distribution function $g(k)$. In general these parameters should be
extracted from experiments. If $E^*$ is identified with spectrum
of physical neutrino, then we are to suggest
the following (probably not complete) list of existing
neutrino experiments, where this spectrum has chance to be found:

a) Neutron beta-decay  (spectrum of electrons near kinematical bound)

b) Tritium beta-decay ($Tr \rightarrow  He^3 + e +\bar{\nu}_e$)
at the end of electron energy spectrum, where at present there exists
discrepancy between theory and experiment.

c) Double neutrinoless beta-decay (test on the Majorano mass).

d) Solar and atmospherical neutrino (oscillations in "vacuum" and matter).

e) Neutrino signal from the supernova (time delay and oscillations)
\vskip 5mm

Authors are deeply grateful to Prs. D.A.Naumov, S.S.Gerstein,
V.A.Matveev and to
Dr. A.E.Kaloshin for the discussion and the critical remarks.
\vskip 5mm

\begin{flushleft}
4. References
\end{flushleft}
\begin{enumerate}
\item
   Y.Nambu, G.Jona-Lasinio, {\it Phys.Rev.} {\bf 122} (1961)
        345; {\it ibid.} {\bf 124} (1961) 246.
\item
   S.P.Klevansky, {\it Rev.Mod.Phys.} {\bf 64}, No.3 (1992)
        649.
\item
   H.Umezawa, H.Matsumoto and M.Tachiki, {\it Thermo-field
   Dynamics and Condensed States}, (North Holland Publishing Company,
   Amsterdam, 1982).
\end{enumerate}
\end{document}